\documentclass{article}

\usepackage{microtype}
\usepackage{graphicx}
\usepackage{subfigure}
\usepackage{booktabs} 

\usepackage{lipsum}
\usepackage{multirow}

\usepackage{hyperref}

\usepackage[accepted]{icml2023}

\usepackage{amsmath}
\usepackage{amssymb}
\usepackage{mathtools}
\usepackage{amsthm}

\usepackage[capitalize,noabbrev]{cleveref}

\theoremstyle{plain}

\theoremstyle{definition}

\theoremstyle{remark}

\usepackage[textsize=tiny]{todonotes}

\icmltitlerunning{Positional Encodings for Light Curve Transformers}

\begin{document}

\twocolumn[
\icmltitle{Positional Encodings for Light Curve Transformers: Playing with Positions and Attention}



\icmlsetsymbol{equal}{*}

\begin{icmlauthorlist}
\icmlauthor{Daniel Moreno-Cartagena}{aaa}
\icmlauthor{Guillermo Cabrera-Vives}{aaa,ccc,ddd,eee}
\icmlauthor{Pavlos Protopapas}{bbb}
\icmlauthor{Cristobal Donoso-Oliva}{ddd,ccc}
\icmlauthor{Manuel Pérez-Carrasco}{ccc,aaa,eee}
\icmlauthor{Martina Cádiz-Leyton}{aaa}
\end{icmlauthorlist}

\icmlaffiliation{aaa}{Department of Computer Science, Universidad de Concepción, Chile} 
\icmlaffiliation{bbb}{John A. Paulson School of Engineering and Applied Science, Harvard University, Cambridge, MA, 02138}
\icmlaffiliation{ccc}{Data Science Unit, Universidad de Concepción, Edmundo Larenas 310, Concepción, Chile}
\icmlaffiliation{ddd}{Millennium Nucleus on Young Exoplanets and their Moons (YEMS), Chile}
\icmlaffiliation{eee}{Millennium Institute of Astrophysics (MAS), Chile}


\icmlcorrespondingauthor{Guillermo Cabrera-Vives}{guillecabrera@inf.udec.cl}

\icmlkeywords{Machine Learning, ICML}

\vskip 0.3in
]



\printAffiliationsAndNotice{\icmlEqualContribution} 

\begin{abstract}
We conducted empirical experiments to assess the transferability of a
light curve transformer to datasets with different cadences and magnitude distributions using various positional encodings (PEs). We proposed a new approach to incorporate the temporal information directly to the output of the last attention layer.  
Our results indicated that using trainable PEs lead to significant improvements in the transformer performances and training times. Our proposed PE on attention can be trained faster than the traditional non-trainable PE transformer while achieving competitive results when transfered to other datasets.


\end{abstract}

\section{Introduction}
\label{sec:Introduction}
The Vera C. Rubin Observatory (LSST; \citealt{ivezic2019lsst}) will produce a vast number of  observations every night in the sky, reaching up to $40$ million events per night where the brightness or location of a source change (\citealt{sanchez2021alert}). The classification of this data is of utmost importance to astronomers as it allows them to acquire information about the physical characteristics and properties of astronomical objects. In recent years, the development of deep learning models has aided in categorization of light curves (\citealt{charnock2017deep, muthukrishna2019rapid, donoso2021effect}). However, light curves pose a considerable challenge as they have different distributions in each of the bands, are irregularly sampled and have varying cadences depending on the telescope at which measurements were taken (\citealt{pasquet2019pelican, yu2021survey}). These peculiarities make it difficult to generate models that are sufficiently generalizable to all astronomical surveys.


Transformers, a type of deep learning model that use self-attention, have demonstrated exceptional performance on light curves (\citealt{pimentel2022deep, donoso2022astromer, astorga2023atat}). In these models, temporal information is conveyed through positional encoding (PE), typically provided by sine and cosine functions with varying frequencies (\citealt{vaswani2017attention}). However, the original proposal of PE was made in the context of text data, assuming uniform word spacing, which differs from the characteristics of astronomical time series. To tackle this challenge, we have focused on enhancing the robustness of the PE definition to effectively generalize to different astronomical surveys.

Currently, there is limited research utilizing positional encodings in transformer models to represent temporal information in light curves. \citet{allam2022paying}, \citet{donoso2022astromer}, and \citet{morvan2022don} employed non-trainable positional encodings by directly inserting timestamps into the predefined function in \citet{vaswani2017attention}, achieving encouraging results compared to traditional deep learning methods. \citet{pimentel2022deep} and \citet{astorga2023atat} proposed a new module based on Fourier decomposition, with $M$ harmonic components and trainable parameters, to induce temporal information. Additionally, \citet{pan2022astroconformer} utilized the rotary positional encoding proposed in \citet{su2021roformer} to explicitly leverage relative positions in the self-attention formulation. However, no studies have analyzed the effect of positional encoding of a transformer model within the light-curve domain.



Motivated by the above, we test different PEs, in a light curve transformer, and evaluate their performance in the reconstruction of astronomical time series and the classification of variable stars. We use the architecture proposed in \citet{donoso2022astromer}, which is a self-supervised light curve transformer model. We perform an empirical comparison and analysis of several PEs on astronomical surveys with different cadences. Additionally, we investigate the potential of a trainable positional encoding and propose a new approach to incorporate the temporal information.

\section{Methods}
\subsection{Baseline light curve transformer model}
\label{subsec:baseline}
Consider light curves of $L$ observations, defined by a vector of magnitudes $x \in \mathbb{R}^{L}$ and times $t \in \mathbb{R}^{L}$ (MJD; Modified Julian Date). A standard transformer add up the projections of the observational and temporal data into a single vector: 
\begin{gather}
    s = FFN(x) + PE(t), \label{eq:InputRepresentationAPE}
\end{gather}
where $FFN(x) \in \mathbb{R}^{L \times d_x}$ represents 
a Feed-Forward Network (FFN)\footnote{Following \citet{donoso2022astromer} we use no activation function for $FNN(x)$.} \comment{is the observational information passed through a Feed-Forward Network}, $d_x$ is the output size of this network, and $PE (t) \in \mathbb{R}^{L \times d_{pe}}$ is the temporal information passed through a positional encoding. To perform the addition operation, $d_{pe}$ must equal to $d_{x}$. As proposed by \citet{donoso2022astromer}, the original positional encoding is expressed as a non-trainable function:
\begin{gather}
    PE(t)_{\ 2j} = \sin\ (\omega_{2j} \cdot t), \label{eq:SinPosASTROMER} \\[3pt]
    PE(t)_{\ 2j+1} = \cos\ (\omega_{2j+1} \cdot t),  \label{eq:CosPosASTROMER} \\[3pt]
    \omega_{j} = \frac{2 \pi}{1000^{\frac{j}{d_{pe}}}}, \label{eq:freqAngularASTROMER}
\end{gather}
\noindent where $\omega_j$ are the angular frequencies, $1000$ defines the lower bound of frequencies, $t$ is the times vector, and $j \in [0,..., d_{pe} - 1]$ are the dimensions of PE with a maximum of $d_{pe}$ frequencies. Note that this PE is a slight modification of the one proposed by \citet{vaswani2017attention}.


The self-attention blocks receive the matrix resulting from the previous step and can be expressed as:
\begin{gather}
    e_{ij}^{(h)} = \frac{s_i W^{(h)}_q \left(s_j W^{(h)}_k\right)^T}{\sqrt{d_k}}, \label{eq:similitud} \\[6pt]
    \alpha_{ij}^{(h)} = \frac{\text{exp}(e_{ij}^{(h)})}{\sum_{l=1}^{L} \text{exp}(e_{il}^{(h)})}, \label{eq:attention} \\[6pt]
    z_{i}^{(h)} = \sum_{j=1}^{L} \alpha_{ij}^{(h)} \left(s_j W^{(h)}_v\right),\label{eq:context}
\end{gather}
%
\noindent where $W^{(h)}_q$, $W^{(h)}_k$, and $W^{(h)}_v$ $\in \mathbb{R}^{d_x \times d_k}$ are trainable weights matrices corresponding to the query ($q$), key ($k$), and value ($v$), respectively. The terms $e_{ij}^{(h)}$, $\alpha_{ij}^{(h)}$, and $z_{i}^{(h)}$ represent the similarity between the query and key vectors, the attention score and the output vector for each observation, respectively. Here, $h \in \{1, \dots ,H\}$ refers to the attention heads, and $d_k$ is a hyperparameter that specifies the embedding size of the self-attention head.  
The output of a self-attention block considers the information of the different heads and is defined as:
\begin{gather}
    z_{i} = \text{Concat} \left(z_{i}^{(1)}, ..., z_{i}^{(H)}\right) W_o, 
\end{gather}
\noindent where $W_o \in \mathbb{R}^{H \cdot d_k \times d}$ is a trainable weight matrix, and $d$ is the dimension of the output $z \in \mathbb{R}^{L \times d}$. This output can be used as input to other multi-head attention blocks to enable the model to capture dependencies at multiple levels of abstraction. The final representation is obtained in the last block. This representation is generated by a decoder FFN that reconstructs the input magnitudes $\hat{x} \in \mathbb{R}^{L}$ in a self-supervised objective task, by minimizing the Root Mean Square Error (RMSE) loss function. The resulting representation can serve as input for subsequent tasks, such as classification or regression. 

\subsection{Positional encodings}
\label{subsec:PE}



\subsubsection{Trainable}
\label{subsubsec:trainable}
In order to capture temporal relationships from the light curves, we replaced the angular frequencies with an embedding layer consisting of $d_{pe}$ trainable parameters. These parameters were initialized with the predefined frequencies given in Eq. (\ref{eq:freqAngularASTROMER}).

\subsubsection{Fourier}
We enhance the learnability and flexibility of the positional representation in Eqs. (\ref{eq:SinPosASTROMER}) and (\ref{eq:CosPosASTROMER}) by modulating it with a Multilayer Perceptron (MLP) \cite{li2021learnable}:
\begin{gather}
\hat{PE}(t) = \left(\Phi_{\text{GeLU}} \left(PE(t) \cdot W_1 + b_1\right)\right) \cdot W_2 + b_2,
\label{eq:PE_mlp}
\end{gather}
where $W_1$ $\in \mathbb{R}^{d_{pe} \times d_m}$ and $W_2$ $\in \mathbb{R}^{d_m \times d_{pe}}$ are trainable parameters, $b_1$ $\in \mathbb{R}^{d_m}$ and $b_2$ $\in \mathbb{R}^{d_{pe}}$ are biases, and $\Phi_{\text{GeLU}}$ is the activation function. Here, $d_m$ is the number of neurons in the hidden layer. In particular, $W_2$ projects the representation to the dimension of the input embeddings. 

\subsubsection{Recurrent}
We followed the approach of \citet{nyborg2022generalized} and used a Gated Recurrent Unit (GRU; \citealt{cho-etal-2014-learning}) to incorporate temporal dependencies at time steps $t$ expressed with the baseline positional encoding shown in Eqs. (\ref{eq:SinPosASTROMER}) and (\ref{eq:CosPosASTROMER}):
\begin{gather}
    o(t) = \text{GRU}\left(PE(t)\right), \\[3pt]
    \hat{PE}(t) = o(t) \cdot W_p + b_p,
    \label{eq:PE_gru}
\end{gather}
where $o(t) \in \mathbb{R}^{L \times d_{pe}}$ is the output of the GRU at each time step, $W_p$ $\in \mathbb{R}^{d_{pe} \times d_{pe}}$ is a trainable weight matrix and $b_p \in \mathbb{R}^{d_{pe}}$ is the trainable bias. 

\subsubsection{Tupe-A}
We follow the approach employed in \citet{ke2020rethinking} for natural language processing (NLP) and separate the mixed correlations produced between observational and temporal information in the attention matrix by redefining Eqs. (\ref{eq:similitud}) and (\ref{eq:context}). To represent the queries and keys of the temporal information expressed by Eqs. (\ref{eq:SinPosASTROMER}) and (\ref{eq:CosPosASTROMER}), we introduce new parameters $U_q, U_k$ $\in \mathbb{R}^{d_{pe} \times d_k}$, respectively:
%
%
%
\begin{gather}
    e_{ij}^{(h)} =   \frac{FFN(x)_i W^{(h)}_q \left(FFN(x)_j W^{(h)}_k\right)^T}{\sqrt{d_k}} \nonumber \\[6pt]
    +\ \frac{PE(t)_i U^{(h)}_q \left(PE(t)_j U^{(h)}_k\right)^T}{\sqrt{d_k}},\\[6pt]
    z_{i}^{(h)} = \sum_{j=1}^{L} \alpha_{ij}^{(h)} (x_j W^{(h)}_v).
\end{gather}
%
For efficiency, we share these new parameters across different multi-head attention blocks \cite{ke2020rethinking}.

\subsubsection{Concat}
Following the same idea and aiming to minimize the noise generated in the attention matrix due to mixed correlations between observational and temporal information, we concatenated them in separate orthogonal spaces:
\begin{gather}
    s = \left[FFN(x)\ ||\ PE(t)\right].
    \label{eq:PE_concat}
\end{gather}
In this case, we utilize the trainable PE described in subsection \ref{subsubsec:trainable}.

\subsubsection{PE on attention (PEA)}
To avoid mixing information, we propose incorporating positional encoding directly into the final representation obtained from the last multi-head attention block:
\begin{gather}
    \hat{z} = z + PE(t),
    \label{eq:PE_PEA}
\end{gather}
where $PE(t)$ is expressed by the baseline positional encoding shown in Eqs. (\ref{eq:SinPosASTROMER}) and (\ref{eq:CosPosASTROMER}) as a non-trainable function. Here, the multi-head attention block takes only the observational information $FFN(x)$ to compute the attention.

\section{Experiments}
\label{sec:Results}

\subsection{Data description}
\label{subsec:Datasets}
For the pretraining stage, we utilized the unlabeled MACHO light curves dataset \citep{alcock2000macho} and excluded curves exhibiting noisy behavior\footnote{We defined noise as points in the light curve with $\lvert Kurtosis \rvert > 10$, $\lvert Skewness \rvert > 1$, and a magnitude error $> 0.1$.}. The dataset comprised a total of $1,529,386$ light curves in the R-band, with a median cadence of $1.00$ days (refer to Appendix \ref{appendix:A} for the cadence distribution). Subsequently, we evaluated the performance of the pretrained transformers on the classification task using a subset of 500 objects per class from the MACHO labeled survey \citep[\emph{Full} hereafter, ][] {cutri2003vizier}. 
To isolate the effect of cadence at this task, we simulated three datasets from the MACHO labeled subset by modifying the cadence of the light curves. Specifically, we removed observations from the light curves at rates of $3/4$, $1/2$, and $1/4$, respectively. At a rate of $3/4$, we removed the last observation out of four, while at a rate of $1/4$, the last three observations were removed. 
To account for external factors, such as changes in the band distribution, we also tested the pretrained transformers on OGLE-III \citep{udalski2004optical}, which contains $358,288$ I-band light curves, and ATLAS \citep{heinze2018first}, which contains $422,630$ orange-band light curves. Specifically, we used a subsample of $500$ objects per class from each labeled data subsets to consider the scenario where we have few labeled data.
The magnitude distribution, cadence distribution and classes of each labeled subset can be found in 
Appendix \ref{appendix:A}.



\subsection{Training details}
\label{subsec:TrainingDetails}
We ran the experiments on a Nvidia RTX A5000 GPU, employing two multi-head attention blocks with $H = 4$ heads and $d_k = 64$ neurons. The model dimensions were set at $d = d_x = d_{pe} = 256$ and $d_m = 64$, with the exception of Concat PE, which employed $d_x = d_{pe} = 128$. \comment{Periods for the PEs where defined using $T = 1,000$.} Light curve windows with a maximum length of $L = 200$ were considered. For light curves that exceeded this length, subsequent time windows were sampled, beginning from a random position. For light curves with fewer than $L$ observations, zero values were padded at the end. Each generated window was subtracted from its observational and temporal mean, creating magnitude and time vectors with zero mean.

For the pretraining stage, we followed the strategy used in \citet{devlin-etal-2019-bert}, masking a percentage of the observations in the light curves. Specifically, we selected $50\%$ of the observations in each light curve for evaluating the reconstruction of the magnitude $\boldsymbol{x}$ in the loss function. Within this percentage, we masked $30\%$ of the observations, replaced $10\%$ with random values, and left the remaining $10\%$ of observations visible. We 
used early stopping with patience of $40$ epochs on the validation loss. The Adam optimizer \citep{kingma2014adam} was used with a learning rate of $10^{-5}$ and a batch size of $2,000$. 

\begin{table*}[t]
\caption{Performance of different positional encodings in the pretraining stage and classification task.}
\label{tab:fixed_RMSE}
\vskip 0.15in
\begin{center}
\begin{small}
\begin{sc}
\begin{tabular}{|l|cc|cccc|cc|}
\toprule
\multirow{3}{*}{PE type} & \multicolumn{2}{c|}{\multirow{2}{*}{MACHO unlab.}} & \multicolumn{4}{c|}{MACHO lab.} & \multirow{2}{*}{OGLE} & \multirow{2}{*}{ATLAS} \\ \cmidrule{4-7}
                         & \multicolumn{2}{c|}{}                       & Full  & $3/4$ & $1/2$ & $1/4$ &                       &                        \\ \cmidrule{2-9} 
                         & RMSE                 & Time (epochs)                & F1 (\%)     & F1 (\%)   & F1 (\%)  & F1 (\%)  & F1 (\%)                   & F1 (\%)                    \\ \midrule

Baseline            &      .170               &       6d 14h (523)               &    71.6 $\pm$ 1.9    &    69.2 $\pm$ 1.9  &   66.2 $\pm$ 1.9   &   63.3 $\pm$ 1.5   &             71.3 $\pm$ 1.1           &         65.8 $\pm$ 1.4              \\ \midrule 
Trainable                &      \textbf{.169}               &       2d 13h (202)               &   72.9 $\pm$ 2.1     &   72.3 $\pm$ 1.0   &  \textbf{71.0} $\pm$ \textbf{1.0}    &   \textbf{69.0} $\pm$ \textbf{0.5}   &           74.9 $\pm$ 1.4             &        65.4 $\pm$ 1.8             \\
Fourier                  &       .170              &    1d 20h (142)                  &   73.0 $\pm$ 1.1    &  70.2 $\pm$ 1.9    &  67.8 $\pm$ 0.9    &  62.9 $\pm$ 2.0    &         72.0 $\pm$ 0.8              &            \textbf{69.6} $\pm$ \textbf{0.1}             \\
Recurrent                &      .197                &           0d 16h (048)          &   67.1 $\pm$ 1.8     &  63.5 $\pm$ 2.5    &  59.7 $\pm$ 1.9    &   54.6 $\pm$ 1.3   &           70.7 $\pm$ 1.1            &          68.3 $\pm$ 0.9              \\
Tupe-A                  &      .219                &      0d 17h (084)               &   67.3 $\pm$ 1.6     &   66.1 $\pm$ 1.4	   &   64.9 $\pm$ 1.0	   &  60.8 $\pm$ 0.9	    &       	71.0 $\pm$ 1.0 	    &      67.5 $\pm$ 0.9                    \\
Concat                   &    .170    &       3d 01h (237)        &    \textbf{73.4} $\pm$ \textbf{1.1}               &          \textbf{73.1} $\pm$ \textbf{1.7}           &    70.9 $\pm$ 1.7     &    \textbf{69.0} $\pm$ \textbf{1.8}   &  74.5 $\pm$ 1.3      &             68.1 $\pm$ 0.6            \\
PEA                   &        .199              &        0d 17h (058)             &   69.7 $\pm$ 0.9     &    68.9 $\pm$ 1.8  &  68.0 $\pm$ 1.0    &  65.5 $\pm$ 2.5    &   \textbf{76.3} $\pm$ \textbf{1.2}                   &         66.9 $\pm$ 1.0                \\
\bottomrule
\end{tabular}
\end{sc}
\end{small}
\end{center}
\vskip -0.1in
\end{table*}

For the classification task, we used two hidden layers of $256$ Long Short-Term Memory (LSTM; \citealt{hochreiter1997long}) units followed by a MLP with a softmax activation function. The dimension of this output layer depends on the number of classes to be classified. We also divided the training and validation sets into $3$ folds with an $80/20$ ratio, respectively. We used early stopping with a patience of $20$ epochs on the validation loss and the Adam optimizer with a learning rate of $10^{-4}$ and a batch size of $512$.

\subsection{Results}
Table \ref{tab:fixed_RMSE} provides the evaluation of a transformer pretrained from scratch using different positional encodings on both simulated and real datasets. We pretrained each transformer on the unlabeled MACHO data and evaluated the reconstruction of the observational information in terms of RMSE and training time. We then used the generated representation for training the classification layers on each labeled dataset and evaluated its performance in terms of F1-score.  Our baseline is the fixed positional encoding described in subsection \ref{subsec:baseline}.

During pretraining, the Trainable PE demonstrated a slight improvement in terms of the reconstruction RMSE and a significant reduction in training time when compared to the baseline. The Fourier and Concat PE matched the performance of the baseline while needing less computational resources. The Fourier PE showed the best performance in terms of both training time and reconstruction performance. Recurrent, Tupe-A, and PEA did not outperform the Baseline in RMSE terms, but converged in less than a day of training. Learning curves are shown in Appendix \ref{appendix:B}.


Since we are analyzing which PE can generate a better representation during the pretraining stage, we keep the transformer, including the PE, fixed when training for the classification task. We start by evaluating the performance of the transformers on the MACHO labeled datasets considering the effect of the change in cadence. The Trainable PE outperformed the Baseline for all cadences. In particular, the degradation of results for sparser light curves is less severe with the Trainable PE than with the non-trainable one. Similarly, Fourier PE performs better than the baseline on three out of four datasets. However, the degradation of results, as evaluated on the $1/2$ and $1/4$ cadences, was similar to the baseline and worse than the Trainable PE. The Recurrent and Tupe-A PE show worse classification performance than the baseline. The Concat PE outperformed the Baseline and achieved three of the four best performances in terms of F1-score. Its degradation was minimal for sparser light curves, and close to the Trainable PE. Finally, PEA outperformed the baseline for cadences of $1/2$ and $1/4$, but did worse for the full and $3/4$ cadences. 

Upon adding changes in the magnitudes distributions using OGLE and ATLAS, we observe that the baseline exhibits inferior overall F1-score performance. The trainable PE model outperforms the baseline in OGLE and demonstrated a similar performance as the baseline for ATLAS. The Fourier PE model showed superior performance in both astronomical surveys with respect to the baseline. However, it did not outperform the Trainable PE model in OGLE. In particular, the Fourier PE model obtained the best F1-score in ATLAS. Similarly, the Recurrent and Tupe-A PE models outperformed the baseline model in ATLAS and while exhibiting a similar performance in OGLE. Finally, the Concat PE and PEA models outperformed the baseline in both OGLE and ATLAS, with the latter obtaining the highest F1-score among all the PE on OGLE.


The separation of temporal and observational information into orthogonal spaces (Concat PE) results in better classification performance on all datasets, yielding the best average F1-score overall. 
Recall that both the Trainable and Concat PE use the same trainable PE: the first add the PE to a vectorized representation of the magnitudes, while the second concatenates these vector. Both of these PEs show a small degradation in results for the MACHO datasets with different cadences, implying that they allow for a better representation of temporal information.

In terms of training time, all the trainable PEs and the proposed PEA exhibit reduced pretraining time compared to the baseline. 
Out of the three models that take less than one day to train (Recurrent, Tupe-A, and PEA), the best classification results for the different MACHO cadence datasets are achieved by our proposed PEA. At the same time, PEA outperforms the Recurrent and Tupe-A PEs when transfered to OGLE, while the three of them achieve similar classification results on the ATLAS dataset (less than 1.6 sigma). This is of particular importance when training large light curve models with massive datasets for next generation surveys such as the LSST.

\section{Conclusion}
\label{sec:Conclusion}

In this work, we have evaluated the transferring potential of a light curve transformer to datasets with different cadences and magnitude distributions.
Our results have demonstrated that using a trainable positional encoding offers advantages over a non-trainable PE baseline, in terms of both model performance and computational efficiency. Additionally, we have highlighted the benefits of separating observational and temporal information within the attention matrix and proposed a new approach for incorporating temporal information directly into the output of the last attention layer. At the same time, our proposed method trains faster than the baseline while achieving competitive classification performances.


\section*{Acknowledgements}
The authors acknowledge support from the National Agency for Research and Development (ANID) grants: FONDECYT regular 1231877 (DMC, GCV, MCL); Millennium Science Initiative Program – NCN2021\_080 (GCV, CDO) and ICN12 009 (GCV, MPC).
%
%


\bibliography{example_paper}
\bibliographystyle{icml2023}

\newpage
\appendix
\onecolumn 
\section{Dataset description.} \label{appendix:A}
\begin{figure*}[!b]
\vskip 0.2in
\begin{center}
\centerline{\includegraphics[width=14cm]{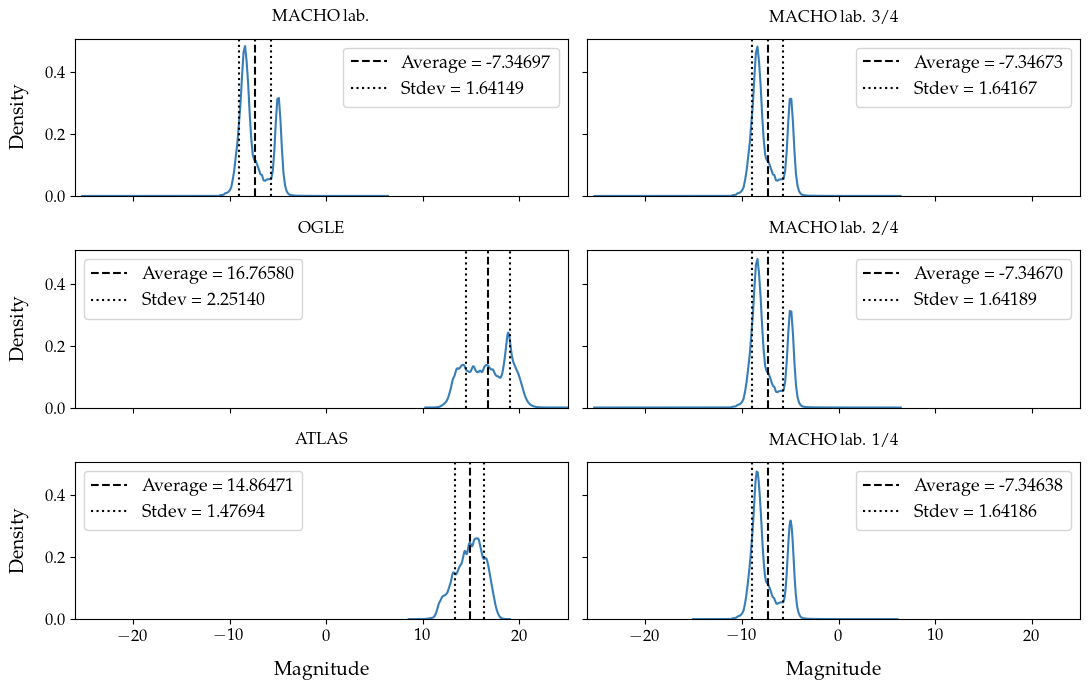}}
\caption{Magnitude distribution of the different data subsets.}
\label{fig:FluxDistribution}
\end{center}
\vskip -0.2in
\end{figure*}
\begin{figure*}[!b]
\vskip 0.2in
\begin{center}
\centerline{\includegraphics[width=11cm]{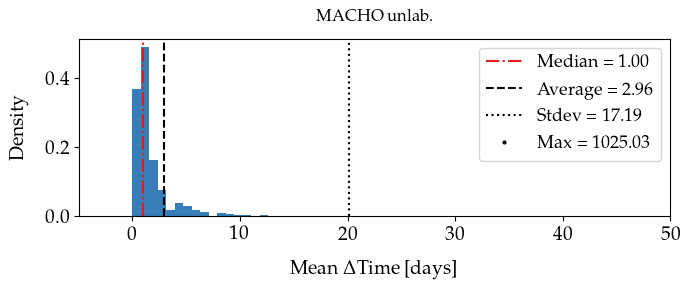}}
\caption{Unlabeled MACHO cadence distribution.}
\label{fig:UnlabeledCadenceDistribution}
\end{center}
\vskip -0.2in
\end{figure*}
\begin{figure*}[!t]
\vskip 0.2in
\begin{center}
\centerline{\includegraphics[width=14cm]{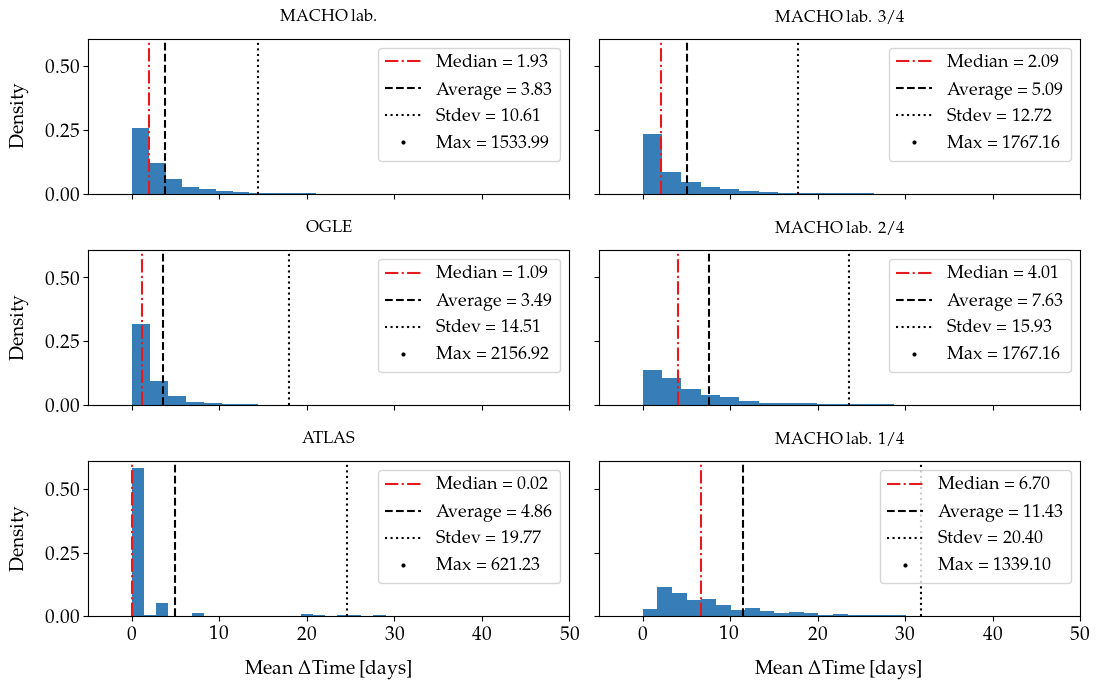}}
\caption{Cadence distribution of the different data subsets.}
\label{fig:LabeledCadenceDistribution}
\end{center}
\vskip -0.2in
\end{figure*}

In this section, we provide essential information to analyze the magnitude, cadence, and class distributions of various astronomical surveys employed in this study. Figure \ref{fig:FluxDistribution} illustrates the magnitude distribution across the different datasets. It shows the similarity in magnitude distributions among the various sets of labeled MACHO (full, $3/4$, $1/2$, and $1/4$) and the disparities in magnitude distributions between the OGLE, ATLAS, and labeled MACHO datasets. Figures \ref{fig:UnlabeledCadenceDistribution} and \ref{fig:LabeledCadenceDistribution} display the cadence distributions for the unlabeled dataset and the labeled datasets, respectively. Key statistical measures are provided to understand the sampling frequency of observations within the light curves. The unlabeled MACHO dataset, which served as the pretraining data for the transformers, exhibits a median of $1.00$ and a mean of $2.96$ with a standard deviation of $17.19$. These values indicate the time gap between successive observations and the temporal separation between groups of observations in the light curves. In particular, the unlabeled MACHO dataset demonstrates a smaller time gap compared to the labeled MACHO dataset, while also exhibiting a higher standard deviation, implying greater temporal separation between groups of observations. The labeled MACHO-derived datasets (3/4, 1/2, and 1/4) exhibit an increase in both the median and standard deviation as light curve observations are reduced. Additionally, they show distinctions compared to both the labeled and unlabeled MACHO datasets. Regarding OGLE, its cadence displays similarities with that of the unlabeled MACHO dataset. However, ATLAS exhibits more pronounced time gaps between groups of observations. The mean is influenced by the standard deviation, while the median indicates that the observations are taken at short time intervals.

Table \ref{tab:labels} displays the classes used for each of the datasets. MACHO labeled has six classes, OGLE has ten classes, and ATLAS has four classes. The modified cadence datasets maintain the same number of classes as the MACHO labeled dataset.
\begin{table*}[h]
\caption{Labels from each of the datasets used in the classification task.}
\label{tab:labels}
\vskip 0.15in
\begin{center}
\begin{small}
\begin{sc}
\begin{tabular}{|l|l|l|l|}
\toprule
Tag    & MACHO lab.        & OGLE                      & ATLAS                  \\ \midrule
EC     & Eclipsing Binary     & Eclipsing Binary          & -                      \\
ED     & -                    & Detached Binary           & Detached Binary        \\
ESD    & -                    & Semi-detached Binary      & -                      \\
Mira   & -                    & Mira                      & Mira                   \\
OSARG  & -                    & Small-amplitude red giant & -                      \\ \midrule
RRab   & RR Lyrae type ab     & RR Lyra type ab           & \multirow{5}{*}{Pulse} \\
RRc    & RR Lyrae type c      & RR Lyrae type c           &                        \\
dsct   & -                    & Delta Scuti               &                        \\ \cmidrule{1-3}
Cep\_0 & Cepheid type I       & \multirow{2}{*}{Cepheid}  &                        \\
Cep\_1 & Cepheid type II      &                           &                        \\ \midrule
SRV    & -                    & Semi-regular variable     & -                      \\
LPV    & Long period variable & -                         & -                      \\
CB     & -                    & -                         & Close Binaries         \\ 
\bottomrule
\end{tabular}
\end{sc}
\end{small}
\end{center}
\vskip -0.1in
\end{table*}

\newpage
\section{Pretraining learning curves.} \label{appendix:B}
\begin{figure*}[h]
\vskip 0.2in
\begin{center}
\centerline{\includegraphics[width=10cm]{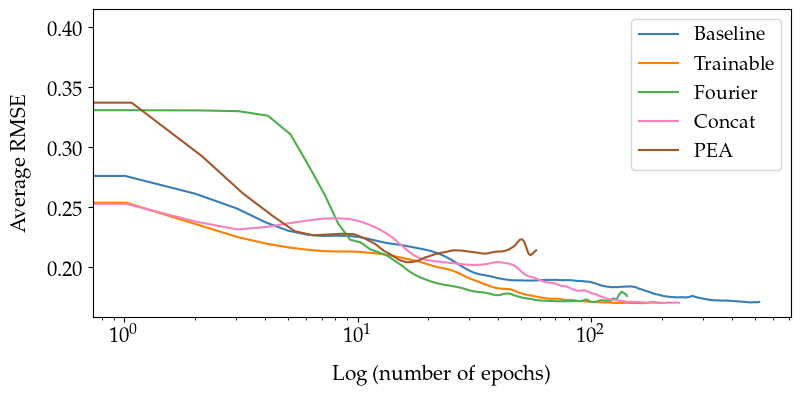}}
\caption{Validation loss in pretraining stage.}
\label{fig:RMSE_pts}
\end{center}
\vskip -0.2in
\end{figure*}

In this section, we present the learning curves on the validation set for the proposed PEA and the positional encodings that achieved superior RMSE during pretraining. Figure \ref{fig:RMSE_pts} illustrates the pretraining of transformers using the same hyperparameters. The y-axis represents the mean value of RMSE with a 4-step window, and the x-axis represents the number of epochs displayed on a logarithmic scale. It is evident that trainable positional encodings such as Trainable, Fourier, and Concat PE achieved comparable RMSE to the baseline with significantly fewer epochs in pretraining ($38.6\%$, $27.2\%$, and $45.3\%$ of baseline epochs, respectively). In contrast, the PEA method initially obtained a higher RMSE than the baseline, but it achieved an average RMSE of $0.204$ earlier than the baseline by utilizing only $57.7\%$ of the epochs.

\end{document}